\documentclass[a4paper,12pt]{article}
\usepackage{latexsym,amsmath,amssymb, amsfonts,mathrsfs}
\usepackage{cite}
\usepackage[arrow,matrix,curve]{xy}
\usepackage{amsmath}

\usepackage{mathrsfs}
\DeclareMathAlphabet{\mathpzc}{OT1}{pzc}{m}{it}

\usepackage{pdfsync}

\usepackage{graphicx}
\usepackage{showidx}
\usepackage{cite}

\usepackage{amssymb}
\usepackage{amscd}
\usepackage{amsfonts}
\usepackage{amsthm}

\numberwithin{equation}{section}

\theoremstyle{definition}
\newtheorem{Theorem}{Theorem}[]

\newtheorem{Proposition}[Theorem]{Proposition}

\theoremstyle{definition}

\newtheorem{Example}[Theorem]{Examples}

\DeclareMathAlphabet{\mathpzc}{OT1}{pzc}{m}{it}

\global\arraycolsep=2pt 
\def\s[#1,#2]{[#1\stackrel{{\displaystyle\star}}{,}#2]}

 1

\def\s[#1,#2]{[#1\stackrel{{\displaystyle\star}}{,}#2]}

\newcommand{\eq}{\begin{equation}}
\newcommand{\eqa}{\begin{eqnarray}}
\newcommand{\en}{\end{equation}}
\newcommand{\ena}{\end{eqnarray}}
\newcommand{\enn}{\nonumber \end{equation}}

\def\sk{\vskip .4cm}
\def\noi{\noindent}

\def\epsi{{\varepsilon}}
\def\st {\star}

\def\D/h{\widehat{\fmslash D}}

\def\al{\alpha}
\def\la{\lambda}
\def\be{\beta}

\def\5bar{{\overline 5}}

\def\FF{\mathcal F}

\def\s'O{\stackrel{_{{\displaystyle\st \footnotesize '}}}{_{^{^{\displaystyle\otimes}}}}}

\def\vphi{\varphi}

\def\D{\Delta}
\def\1s{{1_\st }}
\def\3s{{3_\st }}
\def\2s{{2_\st }}
\def\ef1{{1_\FF}}
\def\ef2{{3_\FF}}
\def\ef3{{2_\FF}}

\def\hbar{\lambda}

\def\cc{\mathbb{C}}

\newcommand{\del}{\partial}
\newcommand{\pa}{\partial}
\newcommand{\eqn}[1]{(\ref{#1})}

\newcommand{\nn}{\nonumber}

\newcommand{\N}{{\cal N}}
\newcommand{\cN}{{\cal N}}
\newcommand{\M}{{\cal M}}
\newcommand{\cM}{{\cal M}}

\newcommand{\A}{{\cal A}}

\newcommand{\matc}{\begin{array}{c}}
\newcommand{\matcc}{\begin{array}{cc}}
\newcommand{\matccc}{\begin{array}{ccc}}
\newcommand{\matcccc}{\begin{array}{cccc}}
\newcommand{\emat}{\end{array}}

\newcommand{\uuu}{u}

\def\bar{\overline}

\newcommand{\HF}{{{\,\,F^{}_{}}^{\!\:\!\!\!\!\!\!\!\!\!{{{\ast}}}~~}}}
\newcommand{\HG}{{{{\,\,G^{}_{}}}^{_{\:}\!\!\!\!\!\!\!\!\!\!\!{{{\ast}}}~~}}}
\newcommand{\HT}{{{_{\,}\,\,T^{}_{}}^{\!\!\;\!\!\!\!\!\!\!\!\!{{{\ast}}}~~}}}
\newcommand{\HbT}{{~\,{\bar T}^{\!\!\!\!\!\!\!\!\!\ast\,~~}}}

\sk\sk
\begin{document}


\begin{titlepage}

{\vspace{-2.8em}}

\hfill {CERN-PH-TH/2013-238}
\sk

\begin{center}
\sk
\sk
{\bf \large Constitutive relations, off shell duality rotations and \\[.4em]the hypergeometric form
  of Born-Infeld theory$^{\dagger}$}

\sk\sk
{\bf Paolo Aschieri,$^{1,2}$ Sergio Ferrara,$^{3,4,*}$ and Stefan Theisen$^5$}
\sk

{\it $^1$Dipartimento di Scienze e Tecnologie
 Avanzate, Universit\`{a} del
 Piemonte Orientale,}\\  {\it $^2$INFN, Sezione di Torino, gruppo collegato di Alessandria }\\
{\it Viale T. Michel 11, 15121 Alessandria, Italy}\\
{\small{\texttt{aschieri@to.infn.it}}}\\[.5em]

        {\it $^3$Physics Department,Theory Unit, CERN, 
        CH 1211, Geneva 23, Switzerland}\\
        {\it $^4$INFN - Laboratori Nazionali di Frascati, 
        Via Enrico Fermi 40,I-00044 Frascati, Italy}\\
{\small\texttt{sergio.ferrara@cern.ch}}\\[.5em]

{\it $^5$Max-Planck-Institut f\"ur Gravitationsphysik,
  Albert-Einstein-Institut, \\14476 Golm, Germany}\\
{\small\texttt{stefan.theisen@aei.mpg.de}}\\[.5em]
\sk
\begin{abstract} 
We review equivalent formulations of nonlinear and higher derivatives theories of
electromagnetism exhibiting electric-magnetic duality rotations
symmetry. We study in particular 
on shell and off shell formulations of this symmetry, at the level of
action functionals as well as of equations of motion.  We prove the conjecture
that the action functional leading to Born-Infeld nonlinear
electromagnetism, that is duality rotation invariant off shell and
that is  known to be a root of an algebraic equation of
fourth order,  is a hypergeometric function. 
\end{abstract}

\setcounter{page}{0}
\end{center}
\sk
$\overline{~~~~~~~~~~~~~~~~~~~~~~~~~~~~~~~~~~~~~~~~~}$

\small{$^\dagger$ Contribution to the proceedings of the conference
BUDS2013, Breaking of supersymmetry and Ultraviolet Divergences in
extended Supergravity, National Laboratories of Frascati,  March 25-28
2013. \\[-.6em]

$^*$ On leave of absence from Department of Physics and Astronomy, University of California Los Angeles, CA 90095- 1547 USA }
 \end{titlepage}

\renewcommand{\thepage}{\arabic{page}}

\section{Introduction}
Electric-magnetic duality is a symmetry of Maxwell electromagnetism
and also, as remarked by Schr\"odinger \cite{Schrodinger},  of the
nonlinear theory of electromagnetism proposed by Born and Infeld \cite{BI}.
This symmetry does not leave the Lagrangians invariant, only the
equations of motion,  and therefore it is not immediately detectable.  
This symmetry was subsequently discovered to be present in extended supergravity
theories \cite{FSZ77,csf,crju}.  In \cite{csf} the first example of
a noncompact duality rotation group was considered, it is due to
scalar fields transforming nonlinearly under duality
rotations. These results triggered
further investigations in the general structure of  self-dual
theories. In particular the symplectic formalism for nonlinear
electromagnetism coupled to scalar and
fermion fields was initiated in \cite{GZ}, there  the
duality  groups were shown to be subgroups of noncompact symplectic
groups (compact groups being recovered in the absence of scalar
fields).  Also nonlinear theories admit noncompact duality symmetry, a
most studied example is  Born-Infeld electrodynamics
coupled to axion and dilaton fields \cite{Gibbons:1995ap}.
A relevant aspect of Born-Infeld theory \cite{BG} is that the spontaneous breaking of $N=2$ rigid
supersymmetry to $N=1$ can lead to a Goldstone vector multiplet whose
action is the supersymmetric and self-dual Born-Infeld action 
\cite{DP, CF}.
Higher supersymmetric Born-Infeld type actions are also self-dual and related to spontaneous
supersymmetry breakings in field theory \cite{KET, KT, KT2, BIK} and in string
theory \cite{KET2, RT}.

Another recent motivation for the renewed study of  duality symmetry
is due to its relevance for investigating the
structure of possible counterterms in extended supergravity. 
After the explicit computations that showed the 3-loop UV finiteness of 
$N=8$ supergravity \cite{Bern}, an explanation based on  
$E_{7(7)}$ duality symmetry was provided 
\cite{Brodel:2009hu, Elvang:2010kc, Beisert:2010jx, Bossard:2010bd}.
Furthermore duality symmetry arguments have also been used to suggest
all loop finiteness of $N=8$ supergravity \cite{Kallosh:2011dp}.
Related to these developments, 
in \cite{BN} a proposal on how to implement 
duality rotation invariant counterterms in a corrected action $S[F]$ leading to a
self-dual theory was put forward under the name of ``deformed 
 twisted self-duality conditions'' . The proposal (renamed ``nonlinear 
twisted self-duality conditions'') was further elaborated in
\cite{CKR} and \cite{CKO}; see also \cite{BCFKR}, and 
\cite{Kuzenko, Kuzenko:2013gr, IZL}, for the supersymmetric extensions
and examples.
The proposal encompasses
theories that depend nonlinearly on the field strength $F$
and also on the partial derivative terms $\partial F, \del\del F, ...\,$. 
That is why we speak of {\it nonlinear and higher derivatives theories}.

The proposal is equivalent to a formulation of self-dual theories using auxiliary fields studied in \cite{IZ2001} and \cite{Ivanov:2003uj} in case of nonlinear electromagnetism without higher derivatives of the
field strength. 
This coincidence has been brought to light in a recent paper
\cite{IZ}.  In \cite{AF} two of us presented a systematic and general study of the different 
formulations of $U(1)$ gauge theories and of self-dual ones. 
This lead to a closed form expression of the duality invariant action functional describing
Born-Infeld theory. 
\sk
Before outlining the content of the present work let us recall the
notion of {\it constitutive relations}. A nonlinear and higher derivative
electromagnetic theory is determined by defining, eventually
implicitly, the relation between the electric field strength $F$
(given by the electric field $\overrightarrow{E}$ and the magnetic induction
$\overrightarrow{B}$ ) and the  magnetic field strength $G$ (given by the magnetic field $\overrightarrow{H}$ 
and the electric displacement $\overrightarrow{D}$). 
We call {\it constitutive relations} the relations defining $G$ in
terms of $F$ or vice versa. Different constitutive relations
determine different $U(1)$ gauge theories.

In this paper we first review and  clarify the relations between
constitutive relations and action functionals in nonlinear and higher derivative
electromagnetism. 
Then we provide a pedagogical analysis of the ``deformed
twisted self duality conditions'' and introduce the action functional
${\cal I}[T^-,\overline{T^-}]$ 
obtained via a Legendre transformation from the usual $S[F]$ action
functional in the field strength $F$.
All theories defined via an action functional $S[F]$ and having duality symmetry have a
formulation via an action functional ${\cal I}[T^-,\overline{T^-}]$  that is off shell
invariant under duality rotations.

We then further study the different formulations of the constitutive relations of nonlinear and higher derivatives electromagnetism and then
of self-dual theories. These different formulations are all equivalent
{\it on shell}.  Finally we prove the conjecture formulated in
\cite{AF} concerning the hypergeometric function expression of the
functional $\cal I$ of Born-Infeld theory. The proof uses
Cauchy residue theorem in order to show that the hypergeometric
function satisfies the algebraic quartic equation characterizing the
functional ${\cal I}$.

\section{U(1) duality rotations in nonlinear and higher derivatives
  electromagnetism \label{dualityrot}}
\subsection{Action functionals from equations of motion\label{AFEOM}}

Nonlinear and higher derivatives electromagnetism is described by the equations of motion
\eqa
&&{\pa}_{\mu}
{\widetilde F}^{\mu\nu}  =0~,\label{max22}\\
&&{\pa}_{\mu}
\widetilde{G}^{\mu\nu}=0~, \label{max11}\\
&&
\widetilde G^{\mu\nu}=h^{\mu\nu}[F,\la]
\label{maxwww}~.
\ena 
The first two simply state that the 2-forms $F$ and $G$ are closed, ${{d}} F={{d}} G=0$, indeed
$\widetilde
F^{\mu\nu}\equiv\frac{1}{2}\epsi^{\mu\nu\rho\sigma}F_{\rho\sigma}$,
$\widetilde
G^{\mu\nu}\equiv\frac{1}{2}\epsi^{\mu\nu\rho\sigma}G_{\rho\sigma}$
(with $\epsi^{0123}=1$). The last set 
$\widetilde G^{\mu\nu}=h^{\mu\nu}[F,\la]$, where $\la$ is the
dimensionful parameter typically present in a nonlinear theory\footnote{Nonlinear and higher derivatives theories of electromagnetism  admit
one (or more) dimensionful coupling constant(s) $\la$.
}, 
are the constitutive relations.  They specify the dynamics and
determine the magnetic field strength $G$ as a functional in terms of the electric field strength $F$, and, vice versa, determine $F$ in term
of $G$, indeed $F$ and $G$ should be treated on equal footing in (\ref{max22})-(\ref{maxwww}).
The square bracket notation  $h^{\mu\nu}[F,\la]$ stems from
the possible dependence of $h^{\mu\nu} $ on derivatives of $F$.

Since in general we consider  curved background metrics
$g_{\mu\nu}$, it is convenient to introduce the $\ast$-Hodge operator;
on an arbitrary antisymmetric tensor $F_{\mu\nu}$ it is defined by 
\eq
\HF{}_{\mu\nu}=
\frac{1}{2\sqrt{g}}
g_{\mu\al}g_{\nu\beta}\,\epsi^{\al\be\rho\sigma}F_{\rho\sigma}
=\frac{1}{\sqrt{g}}\widetilde F_{\mu\nu}~,
\en
where $g=-\det(g_{\mu\nu})$, and it squares to
minus the identity.
The constitutive relations (\ref{maxwww}) implicitly include also
a dependence on the background metric $g_{\mu\nu}$ and for example in
case of usual electromagnetism they read $G_{\mu\nu}=\HF_{\mu\nu}=\frac{1}{\sqrt{g}}\widetilde
F_{\mu\nu}$, while for 
Born-Infeld theory, 
\eq
{S}_{BI}= \frac{1}{\la}\int\!d^4x\,\sqrt{g}\Big( 1-\sqrt{1+\frac{1}{2}\la
  F^2-\frac{1}{16}\la^2(F\HF)^2}\;\Big)~,\label{BILag}
\en
where $F^2=FF=F_{\mu\nu}F^{\mu\nu}$ and 
$F\HF=F_{\mu\nu}\HF^{\mu\nu}$,
they read
\eq\label{BIcr}
{G}_{\mu\nu}=
\frac{\HF{}_{\!\mu\nu}+{1\over 4} \la (F{\HF})\,F_{\mu\nu}}{
\sqrt{1+{1\over 2}\la F^2-\frac{1}{16}\la^2(F{\HF})^2}}~.
\en
The constitutive relations (\ref{maxwww}) define a nonlinear and higher
derivatives extension of electromagnetism because we require that setting $\la=0$ in
(\ref{maxwww})
we recover usual electromagnetism: $G_{\mu\nu}=\HF{}_{\!\mu\nu}$.
\sk
We now recall  \cite{AF} that in the general nonlinear case (where the constitutive relations do not
involve derivatives of $F$) the equations of motion  (\ref{max22})-(\ref{maxwww})
can always be obtained from a variational principle provided they
satisfy the integrability conditions 
\eq\label{intcond}
\frac{\partial
{h}^{\mu\nu}}{\partial F_{\rho\sigma}}=\frac{\partial
{h}^{\rho\sigma}}{\partial F_{\mu\nu}}~.
\en
These conditions are necessary in order to obtain (\ref{maxwww}) from
an action $S[F]=\int \!d^4x \/{\cal L}(F)$. Indeed if\/\footnote{The factor 2 is due to the 
convention $\frac{\partial{F_{\rho\sigma}}}{\partial
  F_{\mu\nu}}=\delta^\mu_\rho\delta^\nu_\sigma\,$ adopted in \cite{GZ}
and in the review \cite{AFZ}. It will be used
throughout the paper.\label{funo}} $h^{\mu\nu}=2\frac{\partial
{\cal L}}{\partial F_{\mu\nu}}$ then (\ref{intcond}) trivially holds.

In order to show that (\ref{intcond}) is also a sufficient condition we recall that the field
strength $F_{\mu\nu}(x)$ locally is a map from spacetime to
$\mathbb{R}^6$ (with coordinates $F_{\mu\nu}$, {\small{$\mu<\nu$}}). We assume 
$h^{\mu\nu}(F,\la)$ to be  well defined functions
on $\mathbb{R}^6$ or more generally on an open submanifold $M\subset
\mathbb{R}^6$ that includes the origin ($F_{\mu\nu}=0$) and that is a star shaped
region w.r.t. the origin (e.g. a 6-dimensional ball or cube
centered in the origin).

Then condition (\ref{intcond}) states that the 1-form
$\mathpzc{h}=h^{\mu\nu}dF_{\mu\nu}$ is closed, and hence, by Poincar\'e lemma,
exact on $M$; we write $\mathpzc{h}=d{\cal L}$. We have ${\cal L}(F)-{\cal L}(0)=\int_\gamma {}_{\!}\mathpzc{h}\,$ for
any curve $\gamma(c)$ of coordinates $\gamma_{\mu\nu}(c)$ such that
$\gamma_{\mu\nu}(0)=0$ and $\gamma_{\mu\nu}(1)=F_{\mu\nu}$. In
particular, choosing the straight line from the origin to the point with
coordinates $F_{\mu\nu}$,
and setting $S=\int d^4x \,{\cal L}(F)$, we immediately conclude:

Under the integrability conditions (\ref{intcond})
locally  the equations of motion of nonlinear electromagnetism 
(\ref{max22})-(\ref{maxwww})  can be obtained  from the action 
\eq
S=\frac{1}{2}\int \!d^4x_{} \int_0^1\! dc \,c F_{\,} \widetilde G_c~,
\en 
where $\widetilde G_c=\frac{1}{c} h(cF,\la)$.

One can also consider the more general case of nonlinear and higher
derivatives electromagnetism. Here too if the theory is obtained from an action functional $S[F]$
then we have 
\eq
S[F]=\frac{1}{2}\int \!d^4x_{} \int_0^1\! dc \, F_{\,} h[cF,\la]~,
\en
that we simply rewrite  $S=\frac{1}{2}\int \!d^4x_{} \int_0^1\! dc \,c
F_{\,} \widetilde G_{c}$.

\sk
\noi{\it Proof. }
Consider the one parameter family of actions
$S_c[F]=\frac{1}{c^2}S[cF]$. 
Deriving with respect to $c$ we obtain 
\eq
-c\frac{\partial S_c}{\partial c}=2S_c-\int \!d^4x ~F\frac{\delta
S_c[F]}{\delta F}~,\label{Ttrace}
\en 
i.e. $-c\frac{\partial S_c}{\partial c}=2S_c-\frac{1}{2}\int \!d^4x
~F\widetilde G_c$. It is easy to see that 
$S_c=\frac{1}{2c^2}\int \!d^4x_{} \int_0^c\! dc' \,c'
F_{\,} \widetilde G_{c'}$ is the primitive with the correct behaviour
under rescaling of $c$ and $F$. We conclude that 
$\frac{1}{c^2}S[cF]= \frac{1}{2c^2}\int \!d^4x_{} \int_0^c\! dc' \,c'
F_{\,} \widetilde G_{c'}$, and setting $c=1$ we complete the proof.

\sk
An equivalent form of the expression  $S=\frac{1}{2}\int \!d^4x_{} \int_0^1\! dc \,c
F_{\,} \widetilde G_{c}$ has been considered, for self-dual theories, in \cite{CKO} and called reconstruction identity.
It has been used to reconstruct the action $S$ from  equations of motion
with duality rotation symmetry in examples with higher derivatives of $F$.

\subsection{Conditions for $U(1)$ duality rotation symmetry of  the
  equations of motion}
Nonlinear and higher derivatives electromagnetism admits $U(1)$ duality rotation symmetry if 
given  a field configuration $F,G$ that satisfies
(\ref{max22})-(\ref{maxwww}) then the rotated configuration
\eq\label{rotFG}
\left(\begin{array}{c}
F' \\
G'
\end{array}\right)=
\left(\begin{array}{cc}
\cos\al & -\sin {\al}\\
\sin\al & \cos\al
\end{array}\right)
\left(\begin{array}{c}
F  \\
G
\end{array}\right)~,
\en
that is trivially a solution of 
${\pa}_{\mu}
{\widetilde F}^{\mu\nu}  =0\,,\;
{\pa}_{\mu}
\widetilde{G}^{\mu\nu}=0\,,
$
satisfies also 
$\widetilde G'_{\mu\nu}=h_{\mu\nu}[F',\la]$, so that $F',G'$ is again a solution of
the  equations of motion.
If we consider an infinitesimal duality rotation, $F\to F+\Delta F$,
$G\to G+\Delta G$ then condition 
$\widetilde G'_{\mu\nu}=h_{\mu\nu}[F',\la]$ reads 
$\Delta\widetilde G_{\mu\nu}=
\int d^4x\, \frac{\delta
  h_{\mu\nu}}{\delta F_{\rho\sigma}}\,\Delta F^{\rho\sigma}$,
i.e., $\widetilde F_{\mu\nu}=-\int d^4x\, \frac{\delta
  h_{\mu\nu}}{\delta F_{\rho\sigma}}\,G^{\rho\sigma}$, that we simply rewrite
\eq\label{basicDR}
\widetilde F_{\mu\nu}=-\int d^4x\, \frac{\delta \widetilde G_{\mu\nu}}{\delta F_{\rho\sigma}}\,G^{\rho\sigma}~.
\en
It is straightforward to check that electromagnetism and Born-Infeld
theory satisfy (\ref{basicDR}).
\sk
If the theory is obtained from an action
functional $S[F]$ (in the field strength $F$ and its derivatives) then
(\ref{maxwww}) is given by
 \eq
\widetilde G^{\mu\nu}= 2\frac{\delta S[F]}{\delta F_{\mu\nu}} ~.\label{Sconst}
\en
In particular it  follows that  
\eq
\frac{\delta{\widetilde G}^{\mu\nu}}{\delta F_{\rho\sigma}}=\frac{\delta
{\widetilde G}^{\rho\sigma}}{\delta F_{\mu\nu}}~,
\en
hence the duality symmetry condition (or self-duality condition)
(\ref{basicDR}) equivalently reads 
$
\widetilde F_{\mu\nu}=-\int d^4x\, \frac{\delta\widetilde  G_{\rho\sigma}}{\delta F_{\mu\nu}}\,G^{\rho\sigma}
$. Now writing $\widetilde F_{\mu\nu}=\frac{\delta}{\delta
  F_{\mu\nu}}\,\frac{1}{2}\!\int \!d^4x \:F_{\rho\sigma}\widetilde F^{\rho\sigma}$ we equivalently have
\eq
\frac{\delta}{\delta F_{\mu\nu}}\int \! d^4x\:(F\widetilde F+G\widetilde G)=0~, \label{NGZ1}
\en
where $F\widetilde F=F_{\rho\sigma}\widetilde F^{\rho\sigma}$ and similarly
for $G\widetilde G$.
We require this condition to hold for any field configuration $F$
(i.e. off shell of (\ref{max22}), (\ref{max11})) and
hence  we obtain the Noether-Gaillard-Zumino (NGZ) self-duality
condition\footnote{Note that (\ref{NGZ2}) (the integrated form of
 the more restrictive self-duality condition  $F\widetilde F+G\widetilde G$) also follows in a straightforward manner  by repeating
  the passages in \cite{GZ} but with $G$ 
  the functional derivatives of the action rather than the partial
  derivatives of the lagrangian \cite{KT, AFZ}. This makes a difference for nonlinear
  theories which also contain terms with derivatives of $F$.}
\eq
\int \! d^4x~(F\widetilde F+G\widetilde G)=0 \label{NGZ2}~.
\en
The vanishing of the integration constant is determined for example by
the condition $G=\HF$  for weak and slowly varying fields,
i.e. by the condition that in this regime the theory is
approximated by usual electromagnetism.

We also observe that  the NGZ self-duality condition (\ref{NGZ2}) is equivalent to
the invariance of $S^{inv}=S-\frac{1}{4}\int \!d^4x\,F\widetilde G$,
indeed under a rotation (\ref{rotFG}) with infinitesimal parameter
$\al$ we have  
$S^{inv}[F']-S^{inv}[F]=-\frac{\al}{4}\int\!d^4x\; (F\widetilde F+
G\widetilde G)=0$.

From this relation it follows that the action $S[F]$ is not invariant under duality
rotations and that under a finite transformation (\ref{rotFG}) we have
\eq
S[F']=S[F]+\frac{1}{8}\int \!d^4x \,\Big(\sin(2\al)(F\widetilde
F-G\widetilde G)-4\sin^2(\al)F\widetilde G\Big)~.
\en
Thus the action changes by the integral of the four-forms $F\wedge
F-G\wedge G$ and $F\wedge G$, that, on the equations of motion
$dF=dG=0$ (cf. (\ref{max22}), (\ref{max11})), are locally  total
derivatives. This is a sufficient condition for the transformation
(\ref{rotFG}) with $\widetilde G^{\mu\nu}=2\frac{\delta S[F]}{\delta F_{\mu\nu}}$
to be a symmetry. 

\sk
We summarize the results thus far obtained: 
The self-duality condition (\ref{NGZ2}) is off shell of (\ref{max22}) and
(\ref{max11}) but on shell of (\ref{maxwww}). The action functional
$S[F]$ provides a variational principle for the equation
(\ref{maxwww}) and under duality rotations changes by a term that on shell of (\ref{max22}) and
(\ref{max11}) is a total derivative

\subsection{Off shell formulation of duality symmetry}\label{offshell}
We here provide an off shell formulation of duality symmetry by
considering a Legendre transformation to new variables. The new action
functional, off shell  of the equations of motion (\ref{max22}),
(\ref{max11}) and  (\ref{maxwww}), is invariant  under duality rotations.
This formulation allows for a classification of duality rotation
symmetric theories (an ackward task using the action functional
$S[F]$). 

An example of functional invariant under duality rotations is provided by the Hamiltonian action functional. Indeed
the Hamiltonian itself (and more generally the energy-momentum tensor)
of duality symmetric theories is invariant under duality
rotations \cite{GZ}\footnote{In a general nonlinear theory the Hamiltonian depends on the magnetic field 
${\overrightarrow B}$ and on the
electric displacement $\overrightarrow{D}=\frac{\delta
  S[F]}{\delta{\overrightarrow{E}}}$, that rotate into each other under the
duality (\ref{rotFG}),
$\Big({}^{~{\overrightarrow B}'}_{-{\overrightarrow D}'}\Big)=\Big({}^{\cos{\al}_{}}_{\sin{\al}^{}}
    {}^{-\sin{\al}_{}}_{~\cos{\al}^{}} \Big)\Big({}^{~{\overrightarrow B}}_{-\\{D}}\Big)$. Since the composite fields
${\overrightarrow{B}}^2+{\overrightarrow{D}}^2$ and $(\overrightarrow{B}\times \overrightarrow{D})^2$ are duality
invariant, Hamiltonians that depend upon these combinations  and their
derivatives are trivially duality invariant and lead to duality
symmetric theories.}. The problem with the Hamiltonian formulation is however the lack of explicit Lorentz
covariance. 

These observations lead to consider a Legendre
transformation of $S[F]$ to an action functional in new variables that
transform linearly under duality rotations and that are Lorentz tensors.

The action $S[F]$ determines the submanifold of equations $\widetilde
G=2 \frac{\del S[F]}{\del F}$ in the plane of coordinates $F$ and
$G$. Equivalently, defining the complex self-dual combinations
\eqa
&&F^-=\frac{1}{2}(F - i\HF)~,~~~~~~~~
\\
&&G^-=\frac{1}{2}(G - i\HG)~,~~~~~~~
\ena
and their complex conjugates $\overline{F^-}=F^+=\frac{1}{2}(\overline
F +i\HF), \overline{G^-}=G^+=\frac{1}{2}(G +
i\HG)$,
the action $S[F^-,\overline{F^-}]=S[F]$ determines the submanifold of
  equations $G^-=-2i\frac{\del S}{\del F^-}$ in the plane of
  coordinates $F^-$, $G^-$.

We want to retrieve this submanifold using  the new variables 
\eqa
&~~~~~~~&T^-=F^--iG^-~,~~\\
&~~~~~~~&\overline{T^+}=F^-+iG^-=2F^--T^-~,~~
~~~~~~\label{OTPM}
\ena
and their complex conjugates 
$\overline{T^-} =F^+
+iG^+$, ${T^+} =F^+
-iG^+=2F^+-\overline{T^-}$. These variables transform simply with a
phase under duality rotations, ${T^-\,}'=e^{i\al}T^-$,
${\overline{T^+}\,}'=e^{-i\al}\overline{T^+}$; hence the
formulation of a theory symmetric under duality rotations should be
facilitated in these variables.
The change of variables $(F^-,G^-)\to (T^-,\overline{T^+})$ is
achieved by first changing from $G^-$ to $T^-$, then by a Legendre
transformation so that $T^-$ become the independent variables and $F^-$ the dependent ones, and finally
changing further the dependent variables from $F^-$ to $\overline{T^+}=2F^--i{T^-}$.
Schematically we undergo the following chain of change of variables 
\eq
(F^- ,G^-)\longrightarrow (F^-,T^-)\longrightarrow
(T^-,F^-)\longrightarrow (T^-, \overline{T^+})~.
\en
More explicitly the equation  in the
$(F^-,G^-)$-plane 
\eq
G^-=-2i\frac{\del S}{\del F^-}
\en
is equivalent to the equation  in the $(F^-,T^-)$-plane
\eq
T^-=\frac{\del  U}{\del F^-}
\en
where
$U[F^-,F^+]=-2S[F^-,F^+]+\frac{1}{2}\int\! d^4x\sqrt{g}  \big({F^-}^2
+{F^+}^2\big)$.
Furthermore, via Legendre transform, this last equation is equivalent to 
the equation in the $(T^-,F^-)$-plane
\eq
F^-=\frac{\delta{V}}{\delta {T^-}}
\en
where
$
V[T^-,\overline{T^-}]=-U[F^-,{F^+}]+\int\!d^4x\sqrt{g}~\,(T^-F^-+\overline{T^-}F^+)~.
$
Finally we rewite this equation in the $(T^-,\overline{T^+})$-plane
as
\eq
\overline{T^+}=\frac{\delta{\cal I}}{\delta{T^-}} \label{Iconst}
\en
where 
\eq
{\cal
   I}[T^-,\overline{T^-}]=2V[T^-,\overline{T^-}]-\frac{1}{2}\int\!d^4x\sqrt{g}~\big({T^-}^2+{\overline{T^-}}^2\big)~.
\en
In conclusion, as pioneered in \cite{Ivanov:2003uj} (in the case of no
derivatives of $F$ in the action), we have that ${\cal
  I}[T^-,\overline{T^-}]$ and $S[F]$ are related by 
\eqa\label{LegendreT}
\!\frac{1}{4}{\cal I}[T^-,&&\!\!\!\!\!\!\!\!\!\overline{T^-}]=S[F]\\ &&\!\!\!+\int\!d^4x\sqrt{g}\,\Big(
 \frac{1}{2}T^-F^--\frac{1}{8}{T^-}^2-\frac{1}{4}{F^-}^2 +
\frac{1}{2}\overline{T^-} F^+-\frac{1}{8}{\overline{T^-}}^2-\frac{1}{4}{F^+}^2\Big)~.\nn
\ena
The equations of motion (\ref{Iconst}) were studied in \cite{BN},
where a nontrivial example of a self-dual action with an infinite
number of derivatives of the field strength $F$ is considered (see
also the generalizations in the appendix of \cite{AF}).
\sk
\sk
Let's now study duality rotations.
We consider $F$ to be the elementary fields and let
$S[F]$ be the action functional of a self-dual theory.
Under infinitesimal duality rotations  (\ref{rotFG}), $F\to F+\Delta F=F-\al G$,
$G\to G+\Delta G=G+\al F$ we have (since
$T^-=F^--\frac{2}{\sqrt{g}}\frac{\delta S}{\delta F^-}$) that $T^-\to
T^-+\Delta T^-=T^--i\al T^-$. We calculate the variation of  (\ref{LegendreT}) under duality
rotations. After a little algebra we
see that 
\eqa\label{sdequiv}
\Delta {\cal I}&=& {\cal I}[T^-+\Delta
T^-,\overline{T^-}+\Delta\overline{T^-}]- {\cal I}[T^-,\overline{T^-}]\\
&=&S[F+\Delta F]-S[F]+\frac{\al}{4}\int\!d^4x\sqrt{g}~\big(G\widetilde 
G-F\widetilde F\big) \nn\\
&=&
-\frac{\al}{4}\int\!d^4x\sqrt{g}~\big(G\widetilde
G+F\widetilde F\big)
=0\nn\ena
where we used that $S[F+\Delta F]-S[F]=\int\!d^4x\; \frac{\delta S}{\delta F}\Delta
F=-\frac{\al}{2}\int\!d^4x\;\widetilde G G$, and the  self-duality conditions (\ref{NGZ2}). 
Hence $\cal I$ 
is invariant under duality rotations.
\sk

Vice versa, we can consider $T^-$, $\overline{T^-}$ to be the elementary
fields and assume  ${\cal I}[T^-,\overline{T^-}]$ to be duality
invariant. Then from 
$2F^--T^-=\frac{1}{\sqrt{g}}\frac{\delta {\cal
    I}[T^-,{\overline{T^-}}]}{\delta T^-_{\;\mu\nu}}$, 
and $F^--iG^-=T^-$, it follows that
under the infinitesimal rotation $T^-\to T^-+\Delta T^-=T^--i\al T^-$ we have
 $F\to F+\Delta F=F-\al G$,
$G\to G+\Delta G=G+\al F$, and from (\ref{sdequiv}) we
 recover the self-duality conditions
(\ref{NGZ2}) for the action $S[F]$.
\sk
This shows the equivalence betweeen the $S[F]$ and the ${\cal
  I}[T^-,\overline{T^-}]$
formulations of self-dual constitutive relations. Hence
the deformed twisted self-duality condition
proposal originated in the context of supergravity counterterms is
actually the general framework needed to discuss self-dual theories
starting from a variational principle.
\sk

We stress that while we needed to use the equations of motion in order to
verify that the action $S[F]$ leads to a duality rotation symmetric
theory, we do not need to use the equations of motion in order to
verify that the action ${\cal I}[T^-,\overline{T^-}]$ is duality
invariant. In the formulation with the ${\cal I}[T^-,\overline{T^-}]$
action functional duality rotations are an off shell symmetry
provided that ${\cal I}[T^-,\overline{T^-}]$ is invariant under
$T^-\rightarrow e^{i\al}T^-$ and $\overline{T^-}\rightarrow e^{-i\al}\overline{T^-}$.

\section{Constitutive relations without self-duality\label{constitutiverelations}}

\subsection{The  ${\cal N}$ and $\M$ matrices}
More insights in the constitutive relations (\ref{maxwww})  can be obtained if we
restrict our study to the wide subclass that can be written as
\eq
{\HG}{}_{\mu\nu}={\cN_2}_{\,} F_{\mu\nu} +{\cN_1}_{\,} \HF{}_{\!\mu\nu}~,
\label{GNN}
\en
where $\cN_2$ is a real scalar field,
while $\cN_1$ is a real pseudo-scalar field (i.e., it is not invariant
under parity, or, if we are in curved spacetime, it is not invariant under
an orientation reversing coordinate transformation).
 As usual in the literature we set 
\eq
\cN=\cN_1+i\cN_2~.
\en
In nonlinear theories $\cN$ depends on the field strength $F$, and in
higher derivatives theories also on derivatives of $F$, we have
therefore in general a functional dependence $\cN=\cN[F,\la]$.
Furthermore $\cN$ is required to satisfy $\cN\to -i$ in
the limit $\la\to 0$ so that we recover classical electromagnetism
when the coupling constant(s) $\la\to 0$, or otherwise stated, in the
weak and slowly varying field limit, i.e., when we discard 
higher powers of $F$ and derivatives of $F$.
Since
$\cN_2\to -1$ for $\la\to 0$, $\cN_2$, at least for sufficiently weak
and slowly varying fields, is invertible. It follows that the
constitutive relation \eqn{GNN} is equivalent to the more duality symmetric one
\eq\label{FFomMFF}
\left(\begin{array}{c}
\HF \\
\HG
\end{array}\right)=
\left(\begin{array}{cc}
0 & -1\\
1 & 0
\end{array}\right)\,\cM\,
\left(\begin{array}{c} 
F  \\
G
\end{array}\right)
\en
where the matrix $\cM$ is given by
\eq
\M(\N)=
\left(\begin{array}{cc}
1 & -\N_1\\
0 & 1
\end{array}\right)
\left(\begin{array}{cc}
\N_2 & 0  \\
0 & \N_2^{-1}
\end{array}\right)
\left(\begin{array}{cc}
1 & 0\\
- \N_1 & 1
\end{array}\right)
=
\left(\begin{array}{cc}
\N_2 +\N_1 \,\N_2^{-1}\,  \N_1 &~ - \N_1 \,\N_2^{-1}  \\
-\N_2^{-1}\, \N_1 &~ \N_2^{-1}
\end{array}\right)~.~~
\label{M(N)}
\en
The matrix $\M$ is symmetric and sympletic and $\M\to -1$ for $\la\to
0$. Actually any such matrix is of the kind 
(\ref{M(N)}) with $\N_1$ real and $\N_2$ real and negative.

Finally, in order to really treat on equal footing the electric and
magnetic field strengths $F$ and $G$, we should consider
functionals ${N}_1[F,G,\la]$ and ${N}_2[F,G,\la]$ such that the constitutive relations
${\HG}={N_2[F,G,\la]}_{\,} F +{N_1[F,G,\la]}_{{\,}} \HF$
are equivalent to (\ref{GNN}), i.e.,  such that on shell of these relations, ${N}_1[F,G,\la]=\N_1[F,\la]$ and 
${N}_2[F,G,\la]=\N_2[F,\la]$.
Henceforth, with slight abuse of notation,
from now on the $\N$, $\N_1$, $\N_2$ fields in
(\ref{GNN})-(\ref{M(N)}) will  in general be functionals of
both $F$ and $G$.

\sk
We now reverse the argument that led from (\ref{GNN}) to (\ref{FFomMFF}).
 We consider 
constitutive relations of the form
\eq\label{FFomMFFp}
\left(\begin{array}{c}
\HF \\
\HG
\end{array}\right)=
\left(\begin{array}{cc}
0 & -1\\
1 & 0
\end{array}\right)\,\cM[F,G,\la]\,
\left(\begin{array}{c} 
F  \\
G
\end{array}\right)
\en
that treat on equal footing $F$ and $G$, and where $\M=\M[F,G,\la]$ is now
an {\sl arbitrary} real
$2\times 2$ matrix  (with scalar entries $\M_{ij}$).
We require
$\M\to -1$ for $\la\to 0$,
so that we recover classical
electromagnetism when the coupling constant $\la\to 0$.
A priory (\ref{FFomMFFp}) is a set of  12 real equations, twice as
much as those present in the constitutive relations (\ref{GNN}). We want only 6 of
these 12 relations to be independent in order to be able to determine $G$
in terms of  independent fields $F$ (or equivalently $F$ in terms of
independent fields $G$). Only in this case the constitutive relations
are well given. In \cite{AF} we show 
\sk
\begin{Proposition}\label{propos3} The constitutive relations  (\ref{FFomMFFp}) 
with $\M[F,G,\la]|_{\la=0}=-1$ are  well given  if and only if on shell of  (\ref{FFomMFFp})  the matrix 
$\M[F,G,\la]$
is symmetric and symplectic. They are  equivalent to the
constitutive relations (\ref{GNN}) provided that on shell the relation between
the $\cal M$ and $\cal N$ matrices is as in (\ref{M(N)}).
\end{Proposition}
Notice that off shell of (\ref{FFomMFFp}) the matrix $\cal M$ does not need to be
symmetric and symplectic. This is  what happens with Schr\"odinger's formulation of Born-Infeld
theory
(see (\ref{BIconst}) and comments thereafter).

\subsection{Schr\"odinger's variables}
Following Schr\"odinger  \cite{Schrodinger, GZS}  it is fruitful to consider the
complex variables
\eq
T=F-iG~,~~\bar{T}=F+iG~.
\en
The transition from the real to the complex variables is given by the
symplectic and unitary matrix $\A^t$
where
\eq
{\cal A}={1\over \sqrt{2}}\left(
\begin{array}{cc}
1  & 1\\
-i & i
\end{array}\right)~~,~~~~{\cal A}^{-1}={\cal A}^\dagger~. \label{defAAm1}
\en
The equation of motions in these variables read $dT=0$, with constitutive
relations obtained applying the matrix ${\cal A}^t$
to (\ref{FFomMFFp}):
\eq\label{TRT}
\left(\begin{array}{c}
\HT \\
\HbT
\end{array}\right)=
-i\left(\begin{array}{cc}
1 & \,0\\
0 & -1
\end{array}\right)\, {\cal A}^t \M \overline{\cal A}
\left(\begin{array}{c} 
T  \\
\bar T
\end{array}\right)~,
\en
where $ {\cal A}^t \M \overline{\cal A}$, on shell of (\ref{TRT}),  is
 complex symplectic and pseudounitary  w.r.t the metric 
$\big({}^1_0{}^{~0}_{-1}\big)$, i.e. it belongs to $Sp(2,\cc)\cap
U(1,1)=SU(1,1)$. It is also Hermitian and negative definite.
These properties uniquely characterize the matrices $ {\cal A}^t \M \overline{\cal A}$ 
as the  matrices 
\eq
\left(\begin{array}{cc}
 -\sqrt{1+\tau\bar \tau^{}} & \, -i \tau\\
i \bar \tau & -\sqrt{1+\tau\bar \tau^{}}
\end{array}\right)
\en
where $\tau=\tau[T,\bar T]$ is a complex field that depends on $T$,
$\bar T$ and possibly also their derivatives. 
We then see that the constitutive relations (\ref{TRT}) are equivalent
to the equations
\eq\label{TccT}
\HT{}_{\!\mu\nu}=i \sqrt {1+\tau\bar \tau} \,T_{\mu\nu} -\tau\,\bar T_{\!\mu\nu}~.
\en
In conclusion
the most general set of equations in the $T$ variables that is well
defined in the sense that it allows to
express  $G=\frac{i}{2}(T+\bar T)$ in terms of
$F=\frac{1}{2}(T+\bar T)$ as in (\ref{GNN})
(equivalently $F$ in terms of $G$) is equivalent, on shell,  to the
equations (\ref{TccT}) for a given $\tau=\tau[T,\bar T]$.
In this sense
 equations (\ref{TccT}) 
are the most general way of defining constitutive relations of
electromagnetism.
The constitutive relations (\ref{GNN})
 are determined by the  complex function ${\cal N}$ (depending on $F,G$ and their derivatives
 ${\cal N}={\cal N}[F,G]$) the equivalent constitutive relations
 (\ref{TccT}) are determined by 
 the complex function $\tau$
(depending on $T, \bar T$ and their derivatives $\tau=\tau[T,\bar T]$).

\section{Schr\"odinger's approach to self-duality conditions\label{SCH}}
In the previous section we have clarified the structure of the
constitutive relations for an arbitrary nonlinear theory of
electromagnetism. The theory can also be with higher
derivatives of the field strength because the complex
field ${\cal N}$, or equivalently the matrix $\cal M$ in
(\ref{FFomMFFp}) of (pseudo)scalar entries, can depend also on
derivatives of the electric and magnetic field strengths $F$ and $G$.

We now further examine the constitutive relations for theories that satisfy the
NGZ self-duality condition 
\eq
F\widetilde F+G\widetilde G=0~,\en 
 i.e.,
$\overline T\widetilde T=0$,
or equivalently,
\eq
\overline T\HT=0~.\label{TGZ}
\en
We multiply  (\ref{TccT}) by $\HT$ and obtain
\eq
-T^2=i\sqrt{1+\tau\overline{\tau}} \,T\HT\label{T1}
\en
It is convenient to consider modulus and argument of these complex
scalar expressions. Setting
\eq
T^2=|T^2|e^{i\al}
\en
from (\ref{T1}) we have
\eq
T\HT =|T\HT| i e^{i\vphi}
\en
We also contract  (\ref{TccT}) with $^\ast{\overline{T}}{}^{\mu\nu}$ and obtain 
$
-T\overline{T}=-\tau \overline{T\HT} ~
$
that implies 
\eq
|\tau|=\frac{T\overline{T}}{|\overline{T\HT}|}~.\label{4.5}
\en
Use of (\ref{T1}) then gives the moduli relations
\eq
|T^2|^2=|T\HT|^2+(T\overline{T})^2~.
\en
The constitutive relations (\ref{TccT}) can also be rewritten using
the chiral variables $T^\pm = T\pm i \HT$, they read
\eq
T^+_{\mu\nu}=te^{i\vphi}\overline{T^-}_{\mu\nu}\label{TptTm}
\en
where $t=\frac{T\overline{T}}{|T^2|+|T\HT|}$.
In order to obtain the explicit relation between the ratio $|\tau|=T\overline T/|T\HT|$
and $t$ we calculate 
\eq|{T^-}^2|(1-t^2)=\frac{1}{2}(|T^2|+|T\HT|)(1-t^2)
=|T\HT|\label{useful}~,
\en 
multiply this last equality by $|\tau|$ and obtain
\eq
(1-t^2)|\tau|=2t~.\label{trrel}
\en

\begin{Example}\label{EEEEX}
{\sl Linear electromagnetism} ($G=\HF$) corresponds to  $|\tau|=0$.
{\sl Born-Infeld nonlinear theory} satisfies the relations
\eq
\HT{}_{\!\mu\nu}=-\frac{T^2}{T\HT} T_{\mu\nu}-\frac{\la}{8}(T\HT)\,\overline T_{\mu\nu}\label{BIconst}
\en
as remarked by Schr\"odinger \cite{Schrodinger}, see \cite{GZS} for a clear
account in nowadays notations. Comparison with (\ref{TccT}) shows 
that, on shell of (\ref{BIconst}) and (\ref{TGZ}), i.e. using
(\ref{T1}) and (\ref{4.5}),
$\frac{T^2}{T\HT}=i\sqrt{1+\tau\overline\tau}$ and
  $\tau=\frac{\la}{8}T\HT$. Hence
Born-Infeld theory is determined by
\eq
|\tau|=\frac{\la}{8} |T\HT|~.\label{rBI}
\en
Schr\"odinger's formulation of Born-Infeld theory uses the freedom,
dicussed in Proposition \ref{propos3}, of considering a matrix $\cal M$
that off shell of (\ref{FFomMFFp})  is not symmetric and
symplectic. Indeed  the term $\frac{T^2}{T\HT}$ is not pure imaginary
off shell. 
Schr\"odinger's elegant variational
principle formulation of Born-Infeld constitutive relations is also  due to
this freedom.
 Defining the ``Lagrangian''
$\Upsilon(T)=\frac{4\,T^2}{T\,\,\,T_{}^{{}^{\!\!\!\!\!\!\;\!\!\!\ast}~}}$
we have that  (\ref{BIconst}) is equivalent to 
\eq\la\;{\bar
  T}^{\!\!\!\!\!\!\!\!\!\ast\,~~}{}^{\!\mu\nu}=\frac{\partial}{\partial
T_{\mu\nu}}\Upsilon(T)~.\en
\end{Example}

\section{Nonlinear theories without higher derivatives\label{nhd}}
We now consider theories (possibly in curved spacetime) that depend only
on the (pseudo)scalars
$F^2$ and $F\HF$, or ${T^-}^{2\!}$ and $\overline{T^-}^{\,2}$.
 Since the
action functional ${\cal I}[T^-,\overline{T^-}]$ studied in Section
\ref{offshell} and the 
scalar field $t$ defined in (\ref{TptTm}) are duality invariant, and under a duality of angle $\al$ we have the
phase rotation  ${T^-}^2\to e^{2i\al}{T^-}^2$, we conclude that ${\cal
I}$ and $t$ depend only on the modulus of ${T^-}^2$, hence 
${\cal I}={\cal I}[T^-,\overline{T^-}]$ and $t=t[T^-,\overline{T^-}]$ simplify to
\eq
{\cal I}=\frac{1}{\la}\int\!d^4x\sqrt{g}\:{I}(\uuu) ~,~~t=t(\uuu)~,
\en
where $I(\uuu)$ is an adimensional scalar function, and the variable
$\uuu$ is defined by

\eq
\uuu_{\,}\equiv_{\,} 2\la|{T^-}^2|_{\,}=_{\,}\la(|T^2|+|T\HT|)~.
\en

\noi Similarly, the constitutive relations (\ref{Iconst}) simplify to 
\eq
{T^+}^{\mu\nu}=
\frac{1}{\la}\frac{\partial I}{\partial {\overline{T^-}_{\!\!\mu\nu}}}=
\frac{1}{\la}\frac{d I}{d \uuu}_{\,}\frac{\partial \uuu\,}{\partial {\overline{T^-}_{\!\!\mu\nu}}}~,
\en and comparison  with (\ref{TptTm}) leads to
\eq
t={2} \frac{d {I}}{d \uuu}~.\label{Ioft}
\en
[Hint: calculate $\frac{\partial \uuu^2}{\partial
\overline{T^-}_{\!\!\!\mu\nu}}$ and
use ${T^-}^2=|{T^-}^2|e^{i\varphi}\,$].

\subsection{Born-Infeld nonlinear theory}

We determine the scalar field $t=t(\uuu)=2\frac{d I}{d u}$ in case of
Born-Infeld theory. This is doable thanks to Schr\"odinger's
formulation (\ref{BIconst}) of Born-Infeld theory, that explicitly
gives $|\tau|=\frac{\la}{8}|T\HT|$, see (\ref{rBI}). 
Then from (\ref{useful}) we have 
\eq
|\tau|=\frac{1}{16}\uuu(1-t^2)~,
\en
and recalling (\ref{trrel}) we obtain \cite{AF, IZ}
\eq
(1-t^2)^2 \uuu= 32 t~. \label{poleq4}
\en
Now in the limit $\uuu\to 0$, i.e., $\la\to 0$, we see from the
definition of $t$  that 
$t\to 0$. 
The function $t=t(\uuu)$ defining Born-Infeld theory
 is  then given by the unique positive root of the fourth order
polynomial equation (\ref{poleq4}) that has the limit $t\to 0$ for $\la\to 0$.
Explicitly,
\eq
t=\frac{1}{\sqrt{3}}\Big(\sqrt{1+s+s^{-1}} - \sqrt{2-s-s^{-1}
+\frac{24\sqrt{3}}{\uuu\sqrt{1+s+s^{-1}}}}~\,\Big) ~,\label{radical1}
\en
where
\eq
s=\frac{1}{\uuu} \Big(216_{\,} \uuu +12 \sqrt{3}\sqrt{108+\uuu^2}_{\,} \uuu+ \uuu^3\Big)^{\mbox{$\frac{1}{3}$}}~.
\label{radical2}\en

\subsection{The hypergeometric function and its hidden identity}
In \cite{CKR} the action functional $\cal I$ and the function $t(\uuu)$
corresponding to the Born-Infeld action  were found via
an iterative procedure order by order in $\la$ (or equivalently in $\uuu$).  The first
coefficients of the power series expansion of $t(\uuu)$ were recognized to
be those of a generalized hypergeometric function, leading to the conclusion
\eqa
t(\uuu)&=&\frac{\uuu}{32} {\,}_3F_2\Big(\frac{1}{2}, \frac{3}{4}, \frac{5}{4};\,
  \frac{4}{3}, \frac{5}{3};\,-\frac{ \uuu^2}{3^3\cdot
    2^2}\Big)~,\label{3F2}\\[.4em]
&=&\frac{2u}{32}\sum_{k=0}^\infty\frac{(4k+1)!}{(3k+2)!k!}\Big(-\frac{\uuu^2}{4^5}\Big)^k\nn
\ena
and, integrating (\ref{Ioft}),
\eq
{I}(\uuu)={6}\left(1-
 {\,}_3F_2\Big(-\frac{1}{2}, -\frac{1}{4}, \frac{1}{4},\,
  \frac{1}{3}, \frac{2}{3};\,-\frac{\uuu^2}{3^3\cdot 2^2}\Big)\right)~.
\en
In \cite{AF} we conjectured, and checked up to order $O(\uuu^{1000})$,  that the expansion in power series of $\uuu$  of the closed form expression of
$t(\uuu)$ derived in (\ref{radical1}), (\ref{radical2}) coincides with
the power series expansion in (\ref{3F2}).

We here present a proof by showing that the power series in
(\ref{3F2}) satisfies the quartic equation (\ref{poleq4}).
We consider the generic power series
\eq
t=\sum_{m=1}^\infty a_mu^m~
\en
with the initial condition $t={\cal O}(u)$ for $u\to 0$, and determine
the coefficients $a_m$ so as to satisfy the quartic equation  (\ref{poleq4}).
The initial condition $t={\cal O}(u)$ for $u\to 0$ is compatible with
(\ref{poleq4}), indeed from  (\ref{poleq4}) we see that for $u\to 0$
we have $t=\frac{u}{32}$.

We extend the variables $t$ and $u$ to the complex plane so that use of  Cauchy's residue theorem 
gives 
\eq
a_m={\frac{1}{2\pi i}} \oint_{C_0} tu^{-m-1}du
\en
We next calculate from (\ref{poleq4}) the differential
\eq
du=32d\frac{t}{(1-t^2)^2}=32\frac{1+3t^2}{(1-t^2)^3} dt~,
\en
and observe  that, since for  $u\to 0$, $t={\cal O}(u)$,
infinitesimal closed paths  surrounding the origin of the complex
$u$-plane are mapped to
infinitesimal ones surrounding the origin of the  complex $t$-plane (that we still
denote $C_0$). We hence obtain
\eqa
a_m&=&\frac{32}{2\pi i}\oint_{C_0} \frac{t+3t^3}{(1-t^2)^3}\frac{(1-t^2)^{2m+2}}{(32t)^{m+1}}dt\nn\\
&=&\frac{1}{32^m \,2\pi i\, }\oint_{C_0} (t^{-m}+3t^{2-m})(1-t^2)^{2m-1}dt\nn\\
&=&\frac{1}{32^m \,2\pi i }\oint_{C_0}
(t^{-m}+3t^{2-m})\sum_{n=0}^{2m-1}(-1)^nt^{2n}
\Big(\begin{array}{cc}{2m-1}\nn\\{n}
\end{array}\Big)dt\nn\\
&=&\frac{1}{32^m}\sum_{n=0}^{2m-1}(-1)^n\Big(\begin{array}{cc}{2m-1}\\{n}
\end{array}\Big)(\delta_{2n-m+1,0} +3\delta_{2n-m+3,0})~.
\ena
We see that only the coefficients $a_m$ with $m$ odd are nonvanishing,
setting $m=2k+1$ we have
\eqa
a_{2k+1}&=&\frac{(-1)^k}{32^{2k+1}}\Big[\Big(\begin{array}{cc}{4k+1}\nn\\{k}
\end{array}\Big) -3 \Big(\begin{array}{cc}{4k+1}\\{k-1}
\end{array}\Big)\Big]\\
&=&(-1)^k\frac{2}{32^{2k+1}}\frac{(4k+1)!}{(3k+2)!k!}~~
\ena 
that proves the conjecture. 

\sk
 As a corollary we have that the hypergeometric function in
 (\ref{3F2})
\eq
{\mathfrak F}(\uuu^2)\equiv{}_3F_2\Big(\frac{1}{2}, \frac{3}{4}, \frac{5}{4};\,
  \frac{4}{3}, \frac{5}{3};\,-\frac{\uuu^2}{3^3\cdot 2^2}\Big)=
{2}\sum_{k=0}^\infty\frac{(4k+1)!}{(3k+2)!k!}\Big(-\frac{\uuu^2}{4^5}\Big)^k
\en
has the closed form expression ${\mathfrak F}(\uuu^2)=\frac{32}{\uuu}t(\uuu)$ where $t(\uuu)$
is given in (\ref{radical1}), (\ref{radical2}), and, because of
(\ref{poleq4}), that it satisfies the ``hidden'' identity
\eq
{\mathfrak F}(\uuu^2)=\Big(1-\frac{\uuu^2}{4^5}{{\mathfrak F}(\uuu^2)}^2\Big)^2~.\label{qeq}
\en

\subsection{General nonlinear theory}
Since Born-Infeld theory is singled out by setting
$|\tau|=\frac{\la}{8}|T\HT|$, and Maxwell theory by setting $|\tau|=0$
(cf. Example \ref{EEEEX}), it is
convenient to describe a general nonlinear theory without higher
derivatives by setting
\eq\label{fuuu}
|\tau|=\frac{\la}{8}|T\HT|f(\uuu)/\uuu
\en
 where $f(\uuu)$ is a positive
function of $\uuu$. We require the theory to reduce to
electromagnetism in the weak field limit, i.e.,
$\HG_{\mu\nu} =-F+ o(F)$ for $F\to 0$. Then we have $T^-={\cal O}(F)$,
$T^+=o(F)$, $u={\cal O}(F^2)$. Hence from (\ref{TptTm}) we obtain $\lim_{u\to 0} t=0$. Moreover from
(\ref{trrel}),  $r={\cal O}(t)$ and from $r=\frac{1}{16}f(u)(1-t^2)$
(that follows from (\ref{fuuu}) and (\ref{useful})\/) $f={\cal
  O}(t)$.
Hence the theory reduces to
electromagnetism in the weak field limit if and only if $\lim_{\uuu\to
  0}f(\uuu)=0$.

From  $r=\frac{1}{16}f(u)(1-t^2)$ (that follows from (\ref{fuuu}) and (\ref{useful})\/)  and (\ref{trrel}) we obtain that the composite function $t(f(\uuu))$ satisfies the fourth order polynomial equation 
\eq
(1-t^2)^2f(\uuu)=32 t~, \label{TFU}
\en
so that
$t(f(\uuu))$ is obtained with the  substitution $\uuu\to f(\uuu)$ in
(\ref{radical1}) and (\ref{radical2}), or in
(\ref{3F2}).

More explicitly, generalizing the results of Example \ref{EEEEX}, we
conclude, as in  \cite{AF}, that the constitutive relations \`a la   Schr\"odinger 
\eq\label{cr5}
\HT{}_{\!\mu\nu}=-\frac{T^2}{T \HT} T_{\mu\nu}-\frac{\la}{8}\frac{f(\uuu)}{\uuu}_{\,}(T\HT)\,\bar T_{\mu\nu}~,
\en
are (on shell)  equivalent to the constitutive relations (deformed twisted
self-duality conditions)
\eq\label{cr6}
{T^+}^{\mu\nu}
=
\frac{1}{2\la} t(f(\uuu)) \,\frac{\partial \uuu}{\partial{\overline{T^-}_{\!\!\mu\nu}}}~,
\en
where $t(f(u))$ satisfies the quartic equation (\ref{TFU}), and we
recall that $\uuu = 2\la|{T^-}^{2}| = \la(|T^2|+|T\HT|)~.$ 
\sk
In other words the appearence of the quartic equation (\ref{TFU})  is a
general feature of the relation between the constitutive relations (\ref{cr5})
and (\ref{cr6}), it appears for any self-dual theory and it is not only a feature
of the Born-Infeld theory.

\sk\sk
\noi
{\large \bf Acknowledgements} \\
P.A. acknowledges the hospitality of  Max-Planck-Institut f\"ur Gravitationsphysik
Albert-Einstein-Institut during commencement of the present work, and the  hospitality of Galileo Galilei
Institute during its completion. The nice
and stimulating atmosphere of the BUDS 2013 conference in Frascati is
also acknowledged.
This work is supported by the ERC Advanced Grant no. 226455, Supersymmetry, Quantum Gravity and Gauge Fields (SUPERFIELDS).


\end{document}